\documentclass[10pt,preprint]{aastex}

\usepackage{amsmath}
\usepackage{amssymb}
\usepackage{amsthm}
\usepackage{graphicx}

\newcommand{\sm}{2M0535$-$05}
\newcommand{\ic}{$I_C$}
\newcommand{\nt}{Paper~I}
\newcommand{\pii}{Paper~II}
\newcommand{\nir}{{\it JHK$_S$}}

\shorttitle{Brown Dwarf Eclipsing Binary}
\shortauthors{G\'omez Maqueo Chew et al.}

\begin{document}

\title{Near-Infrared Light Curves of the Brown Dwarf Eclipsing Binary
2MASS~J05352184--0546085: Can Spots Explain the Temperature Reversal?}

\author{Yilen G\'omez Maqueo Chew and Keivan G. Stassun}
\affil{Department of Physics and Astronomy, Vanderbilt University, Nashville, TN 37235}
\email{yilen.gomez@vanderbilt.edu}

\author{Andrej Pr\v{s}a}
\affil{Department of Astronomy and Astrophysics, Villanova University, Villanova, PA 19085}
\affil{Department of Physics, University of Ljubljana, 1000 Ljubljana, Slovenia}

\and

\author{Robert D. Mathieu}
\affil{Department of Astronomy, University of Wisconsin, Madison, WI 53706}

\begin{abstract}
We present near-infrared \nir\ light curves for the double-lined eclipsing
binary system \object{2MASS J05352184-0546085}, in which both components
have been shown to be brown dwarfs with an age of $\sim 1$ Myr.  We analyze
these light curves together with the previously published \ic-band light
curve and radial velocities to provide refined measurements of the system's
physical parameters.  
The component masses and radii are here determined
with an accuracy of $\sim 6.5$\% and $\sim 1.5$\%, respectively. 
In addition,
we confirm the previous surprising finding that
the primary brown dwarf has a cooler effective temperature than
its lower-mass companion.  Next, we perform a detailed study of
the residual variations in the out-of-eclipse phases of the light curves
to ascertain the properties of any inhomogeneities (e.g.\ spots) on the
surfaces of the brown dwarfs.  Our analysis reveals two 
low-amplitude ($\sim 0.02$ mag) periodic signals,
one attributable to the rotation of the primary with a period of
$3.293 \pm 0.001$~d and the other to the rotation of the secondary with a
period of $14.05 \pm 0.05$~d. Both periods are consistent
with the measured $v \sin i$ and radii.
Finally, we explore the effects on the derived physical parameters of the 
system when spots are included in the modeling of the light curves.
The observed low-amplitude rotational modulations are well fit by cool spots 
covering a small fraction ($\lesssim 10$\%) of the brown dwarfs' surfaces. 
Such small spots negligibly affect the physical properties of 
the brown dwarfs, and thus by themselves cannot explain the primary's 
unexpectedly low surface temperature. To mimic the observed $\sim 200$~K 
suppression of the primary's temperature, our model requires that the primary 
possess a very large spot coverage fraction of $\sim 65$\%. These spots must 
in addition be symmetrically distributed on the primary's surface so as to
not produce photometric variations larger than observed. 
Altogether, a spot configuration in which the primary is
heavily spotted while the secondary is lightly spotted---consistent with the
idea that the primary's magnetic field is much stronger than the 
secondary's---can explain 
the apparent temperature reversal and can bring the temperatures 
of the brown dwarfs into agreement with the predictions of theoretical models.

\end{abstract}


\section{Introduction\label{intro}}

Empirical measurements of the masses, radii, temperatures, and luminosities
of pre--main-sequence (PMS) objects are valuable for the understanding of
star formation.  They delimit the Initial Mass Function, defining the outcome
of star formation and giving the energy scale for the formation process.
They represent the observational tie to the theoretical evolution models
that describe the chronology of stellar evolution, setting the timescales
for circumstellar disk evolution and planet formation.  
In order for these models to accurately describe the physics of PMS evolution, 
they must be tested against observed properties of young stars and brown dwarfs.

For PMS stars with masses lower than 2 M$_{\sun}$, there are
currently only a few tens of objects published with these fundamental 
parameters determined with a precision better than $\sim$10\%
\citep[e.g.,][]{mat07}.
Double-lined eclipsing binary systems allow for a distance independent, 
direct measurement of the masses, radii and, when absolute photometry 
is available, effective temperatures of the components.
Among the techniques for obtaining dynamical masses, the spectro-photometric
modeling of eclipsing binary systems is the only that provides radii for
both components, but more importantly it also renders the most accurate
mass measurements.  If light and radial velocity curves for both
components are available, then the absolute dimensions of the system may 
be obtained \citep[e.g.][]{andersen83}.

The discovery of the system 2MASS J05352184--0546085 (hereafter \sm), 
the first eclipsing binary system
comprised of two brown dwarfs, was presented by \citet{nt06}, hereafter \nt.
With a reported period of $P_o$ = 9.779621 $\pm$ 0.000042 d, \sm\ was found as
part of a photometric survey searching for variability in the Orion Nebula
Cluster.  Through the simultaneous radial velocity and \ic-band light curve
analysis of this fully detached system, 
they obtained masses of $M_1$ = 0.054 $\pm$ 0.005 M$_{\sun}$ and  $M_2$ = 0.034
$\pm$ 0.003 M$_{\sun}$ for the primary and secondary components, respectively, 
with corresponding radii of $R_1$ = 0.669 $\pm$ 0.018 R$_{\sun}$ and  $R_2$ = 0.511 $\pm$
0.026 R$_{\sun}$.  They found a surprising reversal of surface brightnesses
in which the less massive component radiates more per unit surface area 
(i.e.\ has a higher effective temperature) than the more massive one, 
contrary to what is expected for coeval brown dwarfs \citep{bar98}.  

A follow-up analysis of \sm\ 
was presented by \citet{pii} (hereafter \pii) in which it was suggested
that the apparent temperature reversal in \sm\ could be the result of 
preferentially strong magnetic activity on the primary brown dwarf.
This hypothesis was shown by \citet{cha07} to be theoretically plausible,
and was then reinforced empirically when \citet{reiners} found that the
primary brown dwarf rotates $\gtrsim 2\times$ faster and exhibits
$\gtrsim 7\times$ stronger H$\alpha$ emission than the secondary. 
One manifestation of enhanced activity on the primary brown dwarf should
be the presence of large, cool surface spots \citep{cha07}. If present,
such spots should produce photometric variations 
that are periodically modulated by the rotation of the brown dwarf.
Indeed, the presence of low-amplitude variations in the 
\ic-band light curve of \sm\ was noted in \pii,
however an analysis of such variation was deferred to the present paper. 

This paper broadens the previous analyses of \sm\ with the addition of
near-infrared (\nir) light curves, and investigates the intrinsic variability
of the light curves in more detail.  The near-infrared observations and
their reduction are described in Sec.~\ref{data} and analyzed in Sec.~\ref{analysis}.  
A periodicity analysis of the out-of-eclipse phases of the
light curves in Sec.~\ref{period} yields the rotation periods of the two
components of the binary to be $P_{\rm rot,1} = 3.293 \pm 0.001$ d and $P_{\rm rot,2}
= 14.05 \pm 0.05$ d, consistent with the $v\sin i$ measured by \citet{reiners}
and the previously measured radii.
The modeling of the \nir\ light curves together with the previously published
\ic\ light curve and radial
velocity data is described in Sec.~\ref{model}, 
from which we determine refined
measurements of the system's physical parameters.  The apparent temperature
reversal found in the previous studies is confirmed yet again.

Sec.~\ref{spots} incorporates surface spots into the light curve modeling.
In particular, we assess the properties (areal coverage and temperature) of the 
spots that are required to both reproduce the observed low-amplitude 
variations and permit the surrounding photospheric temperatures
of the two brown dwarfs to be in agreement with theoretical expectation for
young brown dwarfs.  
We find that a small cool spot ($\sim 10$\% areal 
coverage and $\sim 10$\% cooler than the surrounding photosphere) on each of the 
brown dwarfs can reproduce the observed low-amplitude variations. Then,
by introducing additional spots that uniformly cover $\sim 65$\% of the 
primary's surface, we are able to simultaneously
reproduce the observed surface brightness ratio of the two brown dwarfs 
(i.e.\ the apparent temperature reversal) while bringing the underlying
temperature of the primary into agreement with the predictions of theoretical models.  
We discuss our findings in
Sec.~\ref{discussion} and summarize our conclusions in Sec.~\ref{summary}.

\section{Near-Infrared Light Curves\label{data}}

This paper focuses primarily on extending the published
spectro-photometric analyses (\nt, \pii) of \sm\ with the addition of the near-infrared 
photometric light curves in the $J$ (1.2 $\mu m$), $H$ (1.6 $\mu m$)
and $K_S$ (2.2 $\mu m$) passbands.  The inclusion of more light curves in
the modeling allows further constraint of the system's parameters, in
particular the temperatures and radii of the components.  The multi-band
analysis also probes the nature of the low-amplitude variability.

\subsection{Near-Infrared Photometric Observations}

The observations of \sm\ presented in this paper were taken in the 2MASS
near-infrared bands \nir\ from October 2003 to April 2006 at Cerro Tololo
Inter-American Observatory in Chile.  They were observed with the SMARTS
1.3-m telescope using the ANDICAM instrument which allows for simultaneous
optical and infrared imaging (the optical measurements have been reported
in \nt\ and \pii).  The observations in the near-infrared were
made in sets of 7 dither positions providing a total of 362 measurements
in $J$, 567 in $H$ and 385 in $K_S$ spread over five observing seasons.
The integration times were
typically of 490 seconds for the \nir\ passbands.  Table~\ref{table-seasons} describes the
observing campaigns in full detail, while Tables~\ref{table-j}--\ref{table-k} provide the
individual measurements in the \nir\ bands.  
The mean near-infrared 
magnitudes of \sm\ are $J=14.646 \pm 0.031$, $H=13.901 \pm 0.043$, and
$K_S = 13.473 \pm 0.031$ \citep{2mass}.

\subsection{Data Reduction}

The data were reduced differently depending on the dome flat acquisition.
For observations made between October 2003 and March 2004, those comprising
the data set I and affecting more than 50 percent of the $H$ light curve,
the dome flats were obtained without information of the mirror's position.
A composite dome flat was created by subtracting a median combination of
$\sim$10 images taken with the dome lights on minus the median combination
of $\sim$10 images taken with the lights off in order to reduce the infrared
contribution in the final images of sources such as the telescope, optical
components and the sky.
The procedure to then reduce data set I consisted of the following steps:
a sky image is formed from the median combination of the 7 dithers; it was
then normalized to the background of each individual image and subtracted
from each separately; every image was then divided by the normalized flat;
the dithers were aligned; the images were cropped, and they were combined
by doing a pixel-by-pixel average.

For images taken from October 2004 onward, dome flats were provided individually
for each of the 7 dither positions, proving essentially helpful in removing
the interference pattern of sky emission lines characteristic to each of
the mirror positions as well as the other infrared contributions.  Each of
the seven furnished dome flats follow the same combination as did the dome
flats described in the previous paragraph.  
The individual dome flats for
each of the mirror's dithers allowed for the creation of separate flats for
each mirror position.  Sky flats were created from the median combination
of $\sim$10 images with slightly different star fields for each distinct
dither position, so that the stars present in the field averaged out and
provided a flat image.  This was possible since the observed field is not a
very crowded one.  For each of the remaining observing seasons, new sky flats 
were created in order to correct for any changes in the dithering and for
any physical changes in the instrument.  The reduction process is slightly
different than for the first data set: the dark was first subtracted
from the raw image; followed by the corresponding normalized sky flat,
which depended on the mirror position at which the images were taken.
The image was then divided by the corresponding normalized dome flat.
Once this was done, the calibration resembles that of data set I: the
dithered images were shifted and cropped in order to be median combined
as to obtain the final image.

\begin{figure}[ht]
\begin{center}
\includegraphics[height=7.5in]{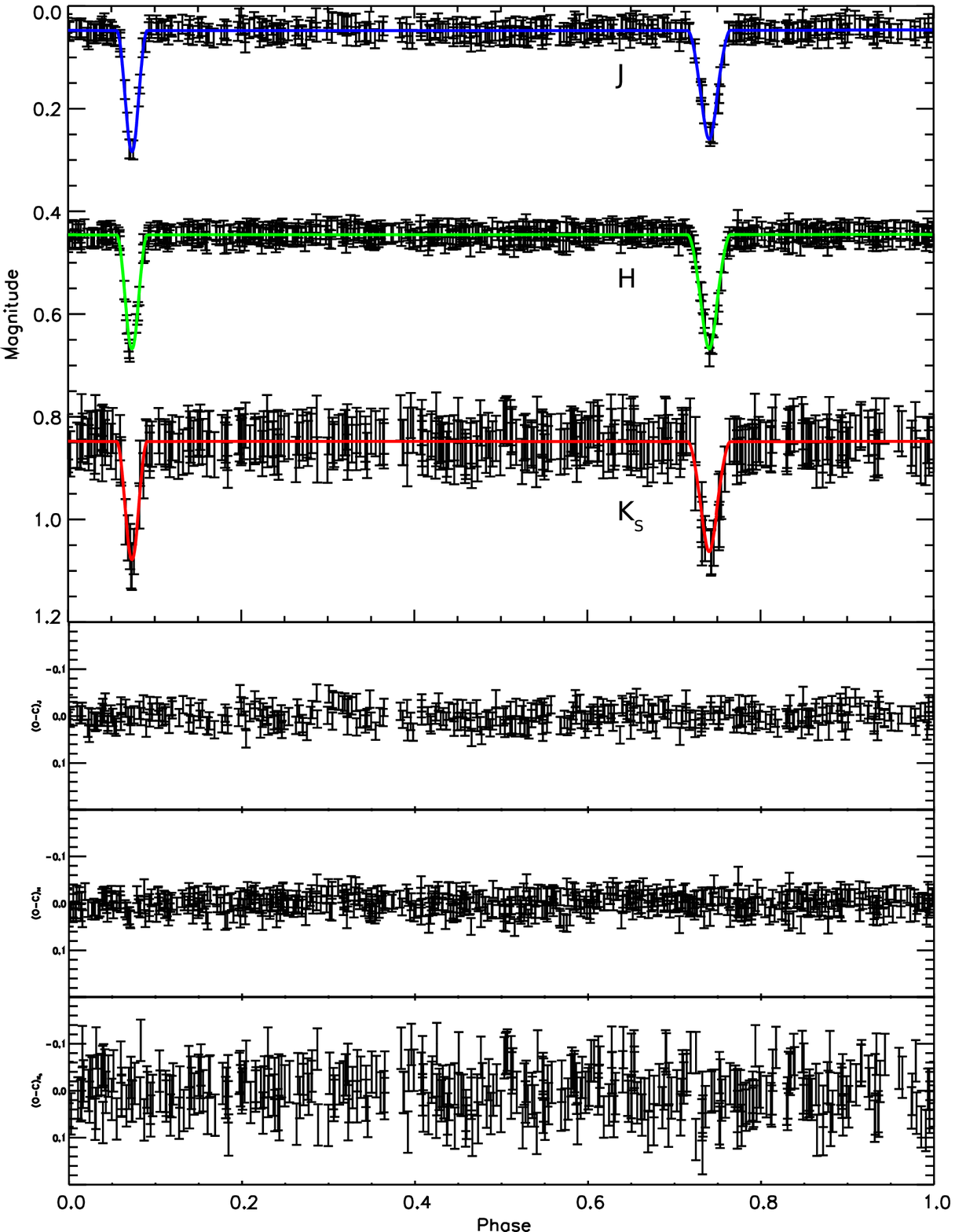}
\caption{\label{fig-jhk-lc}
\nir\ light curves of \sm.  
The observed data points for each band are plotted with their corresponding uncertainties as described on $\S$2.2 and are displaced by 0.7 magnitudes for clarity from the light curve above.  The solid lines represent the light curve model of the simultaneous fit to the radial velocity measurements and the \ic\nir\ photometric data (see \S\ref{model}
for discussion of the modeling procedure, and see
Table~\ref{table-params} for parameters).  The residuals of the fits are also shown.
}
\end{center}
\end{figure}

Differential aperture photometry was done
using the IRAF package APPHOT.  The comparison star 
was chosen because it appears in all of the reduced observations of \sm\ and 
because it is non-variable in the \ic-band observations.  The phased \nir\ light curves
are presented in Fig.~\ref{fig-jhk-lc}.

We are not able to directly measure the absolute photometric precision of the \nir\
light curves because they depend on the assumption that the comparison star is non-variable,
thus we do not report uncertainties on the individual differential photometric measurements
in Tables 2-4. 
However, the standard deviation of the out-of-eclipse portions of the light 
curves gives a measure of the photometric scatter in each of the bands.  
While the $JH$ light curves present a similar scatter, $\sigma_J =
\sigma_H = 0.02$, the interference pattern of the sky emission
lines is more significant in the $K_S$ band making the scatter larger, $\sigma_{K_S} = 0.04$.
As we show below, this photometric scatter includes low-amplitude
intrinsic variations due to the rotation of \sm's components.

\section{Light Curve Analysis\label{analysis}}

The \nir\ light curves described in the previous section are analyzed
for periodicities apart from those due to the eclipsing nature of the binary (\S\ref{period}).
Then they are modeled in conjunction with the available radial velocities
and \ic\ light curve in order to obtain the system's physical parameters (\S\ref{model}).
The thorough treatment of surface spots is introduced to the light curve solution in 
Sec.~\ref{spots}.

\subsection{Rotation periods\label{period}}

The light curves, both in the \ic\ and the \nir\ bands, present
several periodicities.  The most obvious period corresponds to that 
of the eclipses which recur on 
the orbital period, $P_o = 9.779556 \pm 0.000019$~d \citep{pii}.
In addition, the light curves in the observed bandpasses present a low-amplitude 
variability, with a peak-to-peak amplitude of $\sim$0.02--0.04 magnitudes, noticeable 
in the out-of-eclipse phases. 
We speculate that this type of periodic signal is due to the
rotation of one or both components, resulting from spots on their surfaces
rotating in and out of view \citep[e.g.,][]{bouvier,sta99}.  
Another possible explanation for the low-amplitude variations is intrinsic 
pulsation of one or both of the components.  However, young brown dwarfs are 
expected to pulsate with periods of only a few hours \citep{puls} whereas
we find periods of $P_1 = 3.293 \pm 0.001$~d and $P_2 = 14.05 \pm 0.05$~d (see below).
Thus in what follows, for consistency we refer to these periods as 
$P_{\rm rot,1}$ and $P_{\rm rot,2}$.

\begin{figure}[ht]
\includegraphics{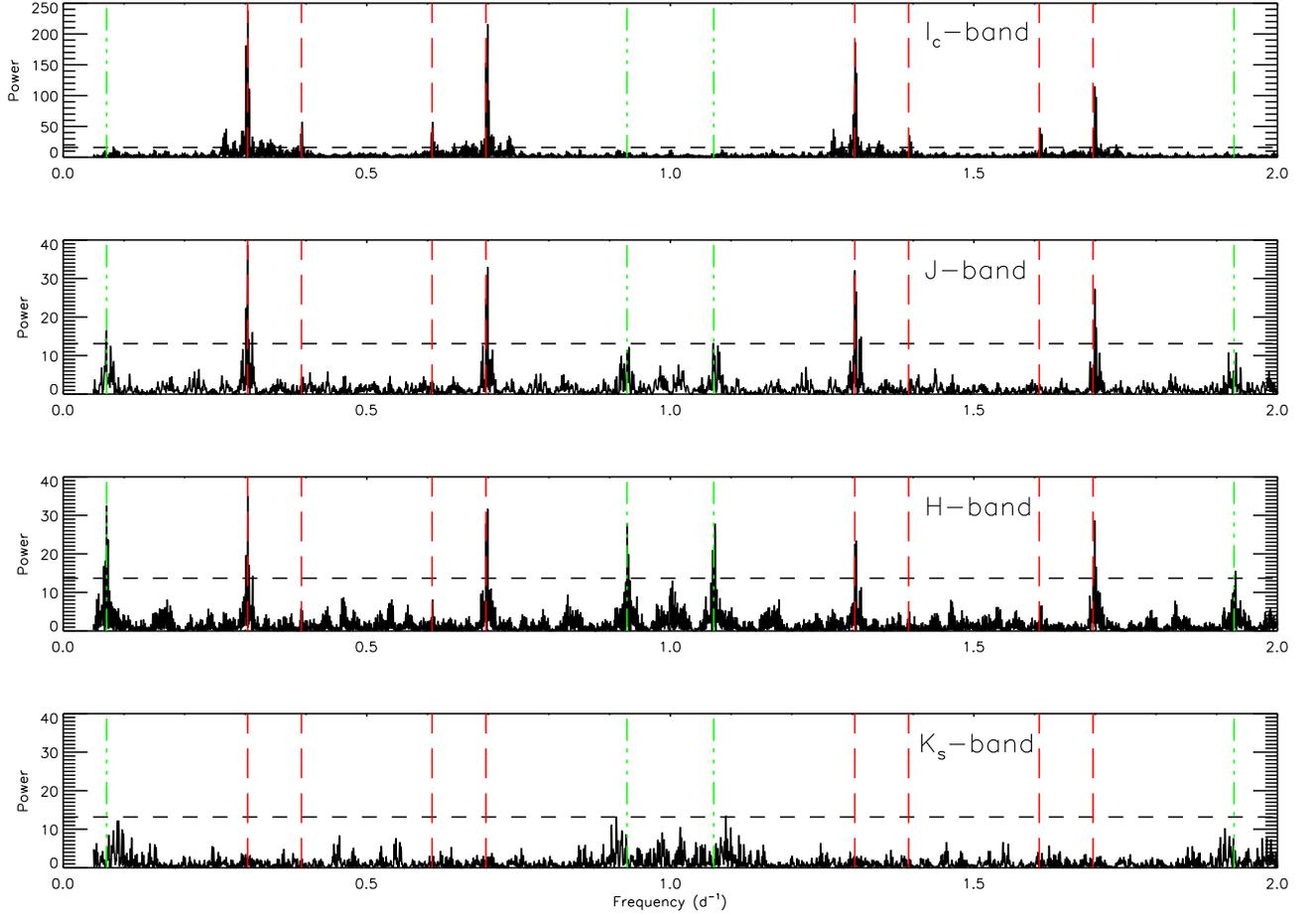}
\caption{\label{fig-periodogram}
Lomb-Scargle Periodograms.  The out-of-eclipse light curves were searched for periodicities using the Lomb-Scargle periodogram
technique \citep{scargle} finding two independent signals with frequencies of $\sim$ 0.30  and $\sim$ 0.07 d$^{-1}$
corresponding to periods of $P_{\rm rot,1} = 3.293 \pm 0.001$ and $P_{\rm rot,2} = 14.05 \pm 0.05$ d respectively.  To assess the
statistical significance of each of the predominant peaks in the power spectrum, the 
false-alarm probability (FAP) was calculated via a Monte Carlo bootstrapping method \citep[e.g.][]{sta99}.  
The horizontal, dashed line denotes the 0.1\% FAP.  
The vertical, long-dashed lines correspond to $P_{\rm rot,1}$ and its 
corresponding aliases and beats; while the vertical, dot-dot-dot-dashed lines correspond to $P_{\rm rot,2}$ and its aliases and beats. 
The out-of-eclipse $I_CJH$ light curves folded over these two identified periods are shown in Fig.~\ref{fig-phased-lc}. }
\end{figure}

The light-curve data in the \ic\ and \nir\ bands corresponding to the 
out-of-eclipse phases were searched for periods between 0.1 and 20 d using
the Lomb-Scargle periodogram \citep{scargle}, well suited for unevenly sampled data. 
The resulting periodograms (Fig.~\ref{fig-periodogram}) show the power 
spectra in frequency units of d$^{-1}$ and present multiple strong peaks.
These represent a 
combination of one or more true independent frequencies together with 
aliases due to the finite data sampling \citep{bkstats}.
The windowing of the data acquisition is of more relevance in the \nir\
bands because a more significant aliasing is produced by including only
data taken through the SMARTS queue observing which has a strong one-day
sampling frequency.

The amplitudes of the periodograms are normalized according to the formulation of \citet{horne} by the total
variance of the data, yielding the appropriate statistical behavior which allows 
for the calculation of the false-alarm probability (FAP).  
The FAP presents the statistical significance of the periodogram  
by describing the probability that a peak of such height would
occur from pure noise.  
To calculate FAPs for the most significant peaks in the
periodogram, a Monte Carlo bootstrapping method \citep[e.g.,][]{sta99} was
applied; it randomizes the differential
magnitudes, keeping the Julian Dates fixed in order to preserve the
statistical characteristics of the data. 
One thousand random combinations
of the out-of-eclipse magnitudes were done with this procedure to obtain the
FAP in each band.  
The resulting 0.1\% FAP level is indicated in the periodograms by the dashed line in
Fig.~\ref{fig-periodogram}.  
Except for the $K_S$ periodogram, multiple peaks are found well above
the 0.1\% FAP level and are therefore highly significant.
The $K_S$ measurements are much noisier than in the $I_C JH$ bands
(Sec.\ \ref{data}), so the lack of significant periodicity in that light curve is
not surprising and we do not consider the $K_S$ light curve further in our
periodicity analysis.

To distinguish the independent periods from their aliases, 
a sinusoid was
fitted to each light curve and subtracted from the data in order 
to filter out the 
periodicity corresponding to the strongest peak in the periodograms.
This peak in the \ic$JH$ bands is that which corresponds to the 3.293 $\pm$ 0.001
d period previously identified in \pii, at a frequency of $\sim$0.30 d$^{-1}$.
This period is not found in the $K_S$ light curve owing to a larger
scatter of the data in that bandpass (\S 2.2).
As expected, the subtraction of the 3.293-d periodic signal removed 
the strongest peak and also its aliases.
The residual light curves were then reanalyzed to identify any additional periods.  

This process revealed another independent frequency at $\sim$0.07~d$^{-1}$ 
which corresponds to a  period of 14.05 $\pm$ 0.05~d. 
This 14.05-d period also manifests itself as  
a three-peaked structure centered at 1~d$^{-1}$ in the $JH$ bands.  
The two exterior peaks of this 
structure have frequencies of 0.93 and 1.07~d$^{-1}$, corresponding to the beat
frequencies between the 14.05-day period and a 1-day period.
The 1-day period is most likely due to the sampling of the observations, 
since the $JH$ bands were observed roughly once per night. 
The \ic\ light curve does not show strong beats against a 1-day period
because this band includes high-cadence data from many observing runs which  
disrupt the 1-day sampling period.
The subsequent filtering of the 14.05-day period, as above, also removes 
its aliases and beats from the periodograms. 

Fig.~\ref{fig-phased-lc} shows the out-of-eclipse light curves of \sm\ 
phased on these two periods,
together with best-fit sinusoids to guide the eye and to quantify the amplitudes
of the variability as a function of wavelength.
Regardless of the order of the filtering, these two independent periods 
were always obtained via this analysis. No other significant periods are found.
We furthermore confirmed that these periods were not present in the light 
curves of the comparison star used for the differential photometry (\S 2.2).

The uncertainty of the periods is given with a confidence interval of
one sigma in the vicinity of the period peaks via the {\it post mortem}
analysis described by \citet{sigmap}.  This method consists of determining
the width of the periodogram's peak at the mean noise power level. The 3.293-d period
has 1-$\sigma$ uncertainties of 0.001-d, 0.003-d and 0.002-d for the \ic-, $J$- and $H$-band respectively;
while for the 14.05-d period the 1-$\sigma$ levels are 0.1-d for the $J$-band and 0.05-d for the $H$-band.

\citet{reiners} reported $v\sin i$ measurements of \sm\
to be $\approx 10$ km s$^{-1}$ for the primary and 
$< 5$ km s$^{-1}$ (upper limit) for the secondary, i.e., the primary rotates 
at least twice as fast as the secondary. 
Moreover, these $v\sin i$ values, together with the radii from \pii\ and
$\sin i \approx 1$, correspond to rotation periods of $\approx 3.3 \pm 0.1$~d and
$> 6$~d for the primary and secondary components, respectively. 
These are consistent with the periods of $3.293 \pm 0.001$~d 
and $14.05 \pm 0.05$~d that we have identified photometrically.

Table~\ref{table-periods} summarizes the appearances of these two periods as a function 
of observing season and passband. 
The 3.29-d period is found consistently in nearly every season of 
observations in all three of the $I_C JH$ filters. 
We fit a sinusoid with a 3.29-d period separately to the data from each of the 
observing seasons and found that while the amplitude of the variability 
remained similar for each, the phase varied from season to season. Evidently,
the 3.29-d period is caused by long-lived features that drift in longitude.
The 14.05-d period is manifested less strongly in the data. While it is
found in the $JH$ light curves in most (but not all) seasons, 
it is detected in only two seasons of the $I_C$-band data. 

Interestingly, while the 3.29-d period manifests an increasing amplitude of 
variability toward shorter wavelengths (Fig.~\ref{fig-phased-lc}, left panels), 
as is expected for spots \citep[either hot or cool; e.g.][]{bouvier},
the amplitude of the 14.05-d periodicity {\it declines} toward shorter wavelengths. 
\citet{maiti} found a similar behavior in the optical variability of the
field L dwarf 2MASSW J0036+1821, and suggested that the photometric variability
in that object is therefore likely caused not by magnetic spots but rather
by dust clouds formed near the surface \citep[e.g.][]{zap05}.
Perhaps the feature on the \sm\ secondary that is responsible for the 
observed 14.05-d period is of similar origin.
Indeed, this would be consistent
with the findings of \citet{reiners} that the \sm\ secondary has a much weaker
magnetic field compared to the primary, and thus may be less likely to produce
strong magnetic spots.

In \S\ref{spots} below, we include spots in our modeling of the \sm\ light curves
in order to demonstrate the 
effects that spots may have on the properties of the magnetically active primary. 
The true physical nature of the inhomogeneity on the magnetically inactive 
secondary does not affect that analysis. For our purposes we emphasize that the 14.05-d
period is consistent with the secondary's measured $v \sin i$ and radius, and thus
we can confidently ascribe that period to the rotation of the secondary.

\subsection{Orbital and Physical Parameters of \sm \label{model}}

Light-curve solutions encompassing the multi-epoch, multi-band photometric
data and radial-velocity measurements were calculated using the software
{\tt PHOEBE} \citep{phoebe} built on top of the 2007 version of the Wilson-Devinney 
algorithm (WD; \citealt*{wd}).
A square root limb-darkening law was adopted,
its coefficients linearly interpolated by {\tt PHOEBE} from the Van
Hamme \citeyearpar{vanh} tables with each iteration. 
Emergent passband intensities are computed based on the passband transmission functions. 

The simultaneous fit of the radial velocities and the \ic\nir\ light curves
was done using the published results from \pii\ as initial parameters for
the modeling.  The first column of Table~\ref{table-params} lists these starting values. The 
solution was then iterated. 
Since we do not have reliable errors on the individual \nir\ measurements
(see \S 2.2), the data points were assigned equal weight and then
the overall weight of each light curves was
set to the inverse-square of the r.m.s.\ of the residuals relative to the fit 
from the previous iteration. 
The primary's temperature is taken to be $T_{\rm eff,1} = 2715 \pm 200$~K,
where the uncertainty is dominated by the systematic uncertainty of
the spectral-type--$T_{\rm eff}$ scale (\pii).
We emphasize that the uncertainty on the individual
component temperatures does not represent the high accuracy with which
the quantities directly involved in the light curve fitting are
determined, namely the ratio of the temperatures.
In addition to setting $T_{\rm eff,1}$ to a fixed value, 
the orbital period $P_o$ was also kept constant.  
The synchronicity parameters are obtained from the rotation periods (\S\ref{period})
such that $F_1 = \omega_{\rm rot,1}/\omega_{\rm orbital} = 2.9725 \pm 0.0009$ and $F_2 =
\omega_{\rm rot,2}/\omega_{\rm orbital} = 0.6985 \pm 0.0025$. 
The free parameters to be obtained from the modeling were: 
the inclination angle $i$, 
the semi-major axis $a$, 
the orbital eccentricity $e$, 
the argument of the periastron $\omega$,
the systemic radial velocity $v_\gamma $, 
the mass ratio $q$ and 
the secondary's surface temperature $T_{\rm eff,2}$, through the determination of the temperature ratio $T_{\rm eff,2}/T_{\rm eff,1}$.
Because the primary's radius is small compared to the semi-major axis 
($R_1 / a = 0.08$), reflection effects are assumed to be negligible 
\citep[reflection effects generally only become important for
$R_1 /a \gtrsim 15$\%; e.g.][]{wil90}. 

A direct output of the Wilson-Devinney algorithm that underlies {\tt PHOEBE} is
the formal statistical errors associated with each of the fit parameters, as
well as a correlation matrix that provides insight into the often complex
interdependencies of the parameters.
In order to explore these parameter correlations and solution degeneracies 
more carefully, and to thus
determine more robust parameter uncertainties, we performed a thorough 
Monte Carlo sampling of the parameter hyperspace using the {\tt PHOEBE} code's
scripting capability.
An examination of the parameter correlation matrix revealed that there are
two particularly strong parameter degeneracies in our dataset: (1) between 
the inclination, $i$, and the surface potentials, $\Omega$; and (2) between the 
temperature ratio, $T_{\rm eff,2}/T_{\rm eff,1}$, and the radius ratio, $R_2/R_1$.

\begin{figure}
\begin{center}
\includegraphics[width=7.0in]{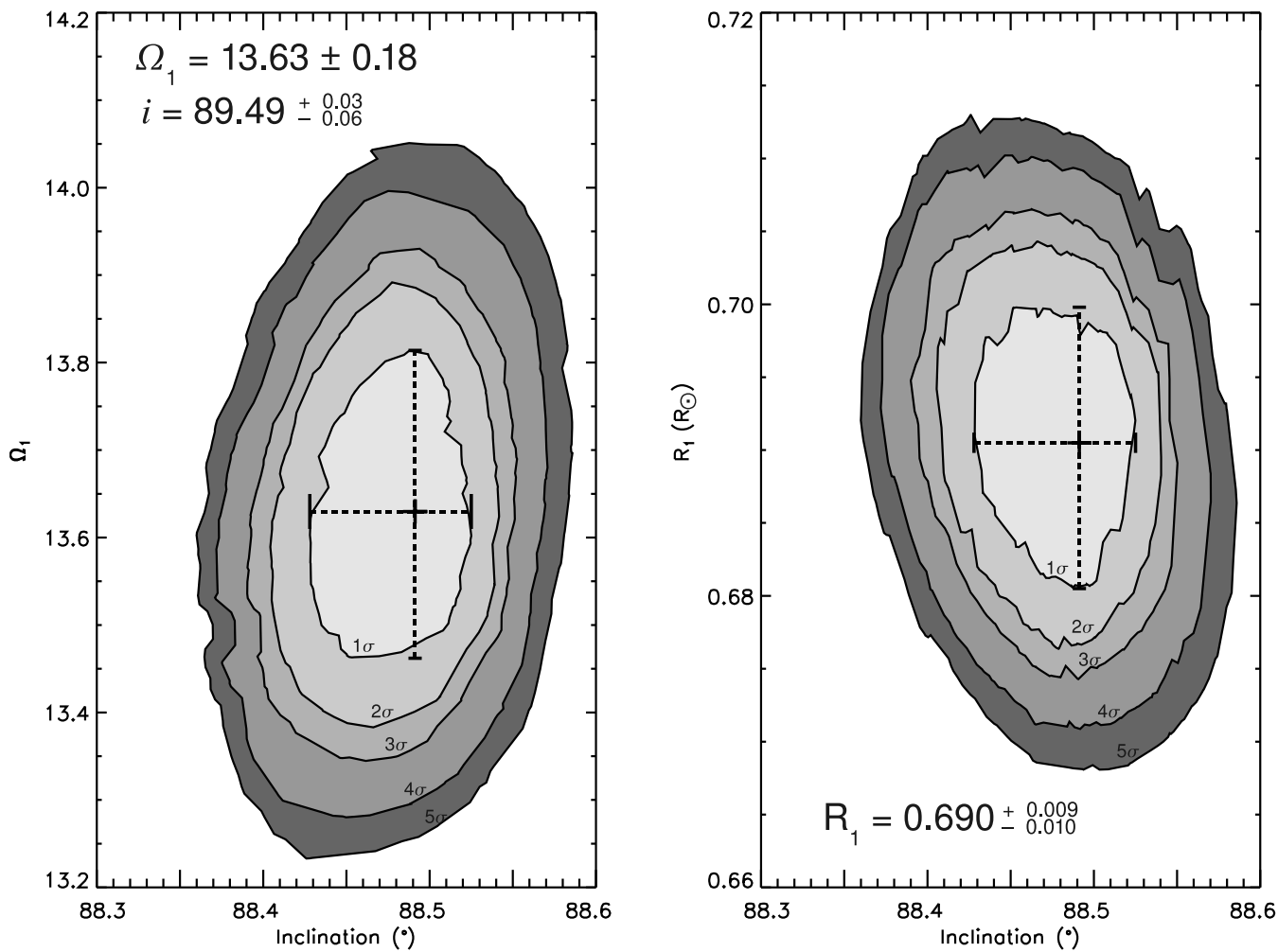}
\caption{\label{fig-iomega}
Joint Confidence Interval between the Inclination Angle $i$ and the Primary Surface Potential $\Omega_1$. The Monte Carlo sampling of this cross section allows for the heuristic errors associated with the available data to be estimated given by the variation of $\chi^2$ with $i$ and $\Omega_1$ (\S\ref{model}).  Because of the intrinsic degeneracy of the binary problem and the data's uncertainties, closely correlated parameters must be explored to ensure that the system's solution falls within the global minimum of the cost function.  The cross represents the point at which the $\chi^2$ of the fit attains its minimimum value and shows the 1-$\sigma$ uncertainties for each of the parameters given by the smallest contour. Each subsequent contour symbolizes an increase of 1 $\sigma$.  The right panel shows the same sampling of the $i - \Omega_1$ cross section in terms of the primary radius.  } 
\end{center}
\end{figure}

Fig.~\ref{fig-iomega} shows the resulting joint confidence interval for 
$i$ and $\Omega_1$ given by the 
variation of $\chi^2$ with these two parameters around the solution's minimum.  
The shaded contours correspond to confidence intervals following a
$\chi^2$ distribution with two degrees of freedom,
with the first contour at the 1-$\sigma$ confidence level and each
subsequent level corresponding to an increment of 1~$\sigma$.
The $i$--$\Omega_1$ cross section was sampled 
by randomly selecting a value for $i$ between 87$^\circ$ and
90$^\circ$, and one for 
$\Omega_1$ between 12.0 and 14.5, 
rendering a more complete coverage of the parameter hyperspace.
We marginalized over the remaining system parameters, notably the 
strongly correlated $\Omega_2$.
This analysis yields uncertainties around the best-fit values of:
$i =  88.49_{-0.06}^{+0.03}$ degrees and $\Omega_1 = 13.63 \pm 0.18$, 
the latter corresponding to a primary radius 
of $R_1 = 0.691_{-0.010}^{+0.009} \ {\rm R}_\sun$.
The secondary's best-fit radius and its uncertainties follow directly through
the ratio of the radii (discussed in the next paragraph).

\begin{figure}
\begin{center}
\includegraphics{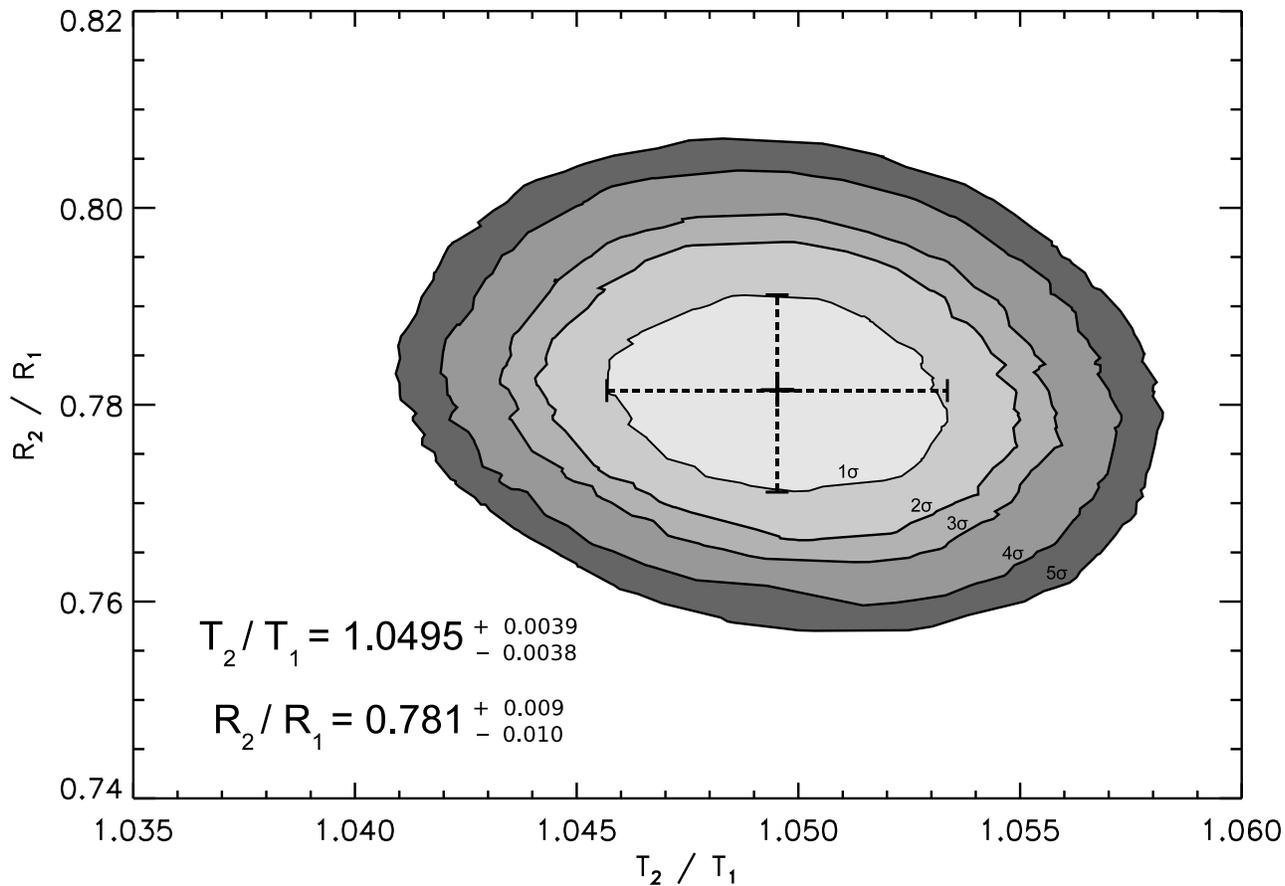}
\caption{\label{fig-tratio}
Joint Confidence Interval between ($T_{\rm eff,2}/T_{\rm eff,1}$) and ($R_2/R_1$).  This parameter hyperspace is of particular interest in the case of \sm \ because of the apparent temperature reversal it presents. Similar to Fig.~\ref{fig-iomega}, the cross represents the point at which the $\chi^2$ of the fit attains its minimimum value and shows the 1-$\sigma$ uncertainties for each of the parameters given by the smallest contour. Each subsequent contour symbolizes an increase of 1 $\sigma$.}
\end{center}
\end{figure}

The $(T_{\rm eff_2}/T_{\rm eff,1})$--$(R_2/R_1)$ plane, 
shown in Fig.~\ref{fig-tratio}, is of 
particular interest because of the apparent temperature reversal that \sm\ presents.  
This parameter cross section was explored keeping the $T_{\rm eff}$ of 
the primary fixed at 2715~K while varying the $T_{\rm eff}$ of the 
secondary between 2700 and 2925~K.
The primary radius was varied randomly 
between 0.635 and 0.758 R$_\sun$, while minimizing for the secondary radius.  
The resulting uncertainties about the best-fit values are:
$T_{\rm eff,2}/T_{\rm eff,1}=1.0495^{+0.0039}_{-0.0038}$
and $R_2/R_1=0.781_{-0.010}^{+0.009}$.  
Note that these errors determined from our Monte Carlo sampling procedure 
are larger than the formal statistical errors by $\sim$50\%.

Finally, we separately performed a fit of the radial velocity data alone for the orbital 
parameters that most directly determine the masses, namely: $a\;sin\;i$, $q$, and $v_\gamma$ 
in order to more conservatively estimate the 
uncertainties in these parameters.
These orbital parameters should not depend on the light curves, 
 however we found that purely statistical correlations between these parameters
 and other system parameters tended to drive down the formal errors in the masses to
 unrealistically small values.   
We include $e$, $\omega$, and the time of periastron in the fit, but 
for these parameters we deferred error estimates to the simultaneous fit to the 
radial velocity and light curve data.  
Therefore we adopted the uncertainties in $a\;sin\;i$, $q$, and
$v_\gamma$ from the radial velocity fit, the uncertainties in $i$, $\Omega_1$, $\Omega_2$,
 and $T_2/T_1$ from the Monte Carlo sampling, and 
the uncertainties of other parameters from the simultaneous fit to 
the radial velocity and light curve data.  We then propagated these uncertainties 
into the final errors of the parameters that depend on these quantities,
such as the masses and radii. 

The final parameters for \sm\ resulting from our joint
analysis of the radial velocities and \ic\nir\ light curves, and with
uncertainties determined as described above, are summarized in the last column 
of Table~\ref{table-params}. The results are in agreement with those previously published,
although the uncertainties in many parameters have now improved compared to those 
reported in \pii. 
For example, the uncertainties in the component masses has decreased 
from $\sim$10\% to $\sim$6.5\%, and the radii from $\sim 5$\% to $\sim$1.5\%. 
This improvement arises primarily because of the improved determination of 
$e$ and $\omega$ through the addition of the \nir\ light curves,
thus improving the determination of the time of periastron passage. 

As in the previous analyses of \sm\ (\nt, \pii), we find again  
a reversal of effective temperatures from what would be expected from 
the observed mass ratio  (i.e.\ the higher mass
primary is cooler than the secondary) at high statistical significance.
This surprising result is now confirmed on the basis of a
full analysis including radial velocities and four light curves (\ic\nir) together.

\section{Surface Spots\label{spots}}

In \S\ref{period} we found clear evidence of two separate low-amplitude 
variations in the light curves of \sm\ with periods of 3.29~d and 14.05~d.
PMS objects are typically found to be photometrically variable
\citep[e.g.,][]{bouvier,carp}, and this variability is in almost all
cases attributable to the presence of magnetic ``spots" (akin to sunspots),
to accretion from a circumstellar disk, or both.
However \sm\ has been shown to not possess circumstellar or circumbinary material
and thus is not currently accreting
\citep{mohanty}. Pulsations have been suggested in a few brown dwarfs,
but are expected to have characteristic periods of only a few hours \citep{puls}.

In this section we explore the effects of surface spots on the light curves for
the purpose of explaining the periodic variations found in \S\ref{period},
and to assess whether such spots might be able to explain the 
surprising reversal of effective temperatures in the system (\S\ref{model}).

\begin{figure}[ht]
\begin{center}
\includegraphics{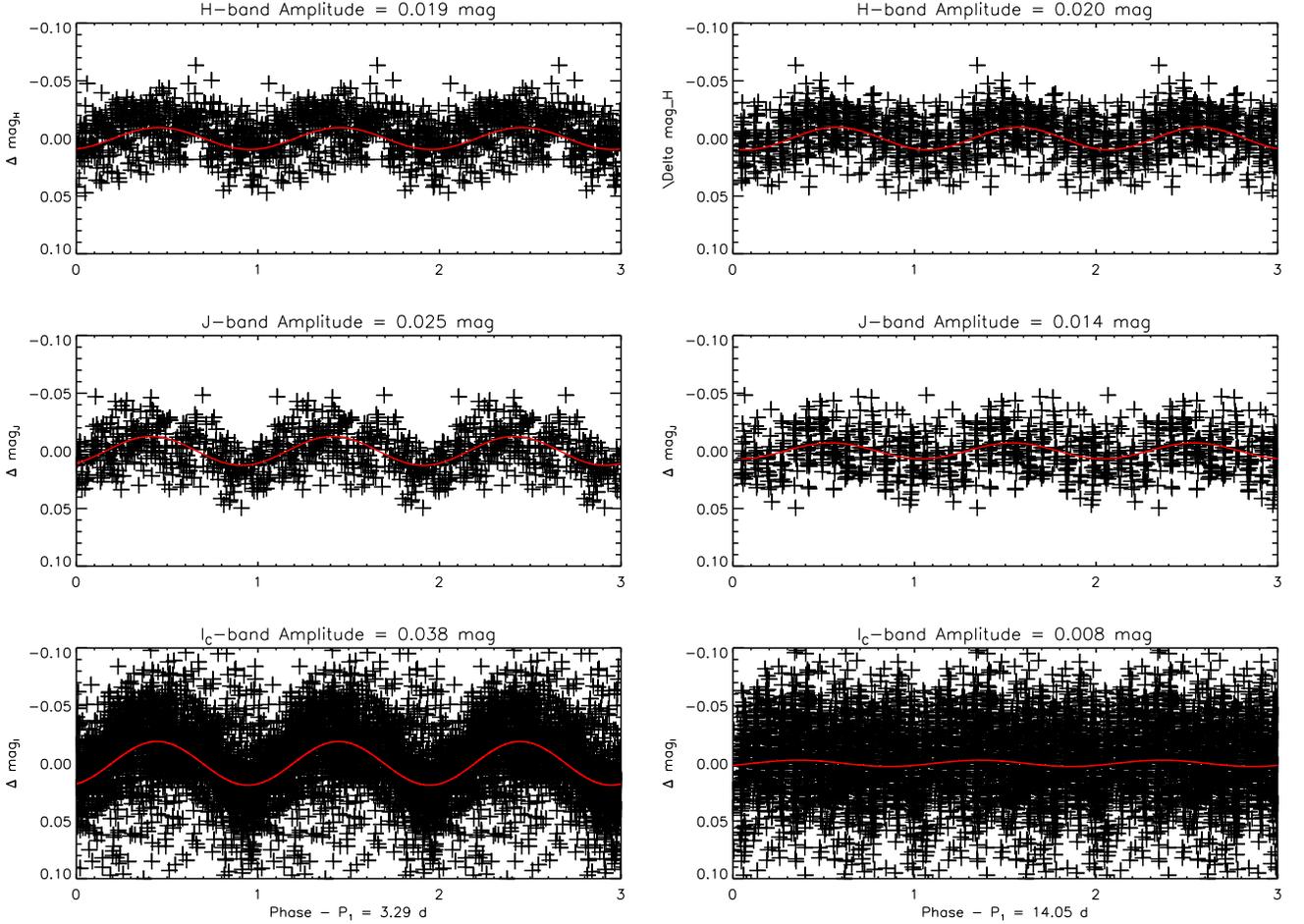}
\caption{\label{fig-phased-lc}
Out-of-Eclipse $I_C JH$-band Light Curves. The low-amplitude photometric variability is made evident by phasing the out-of-eclipse light curves to the two individual periods found from the periodogram analysis (see Fig.~\ref{fig-periodogram}).  The 3.293-d period attributed to the rotation of the more massive brown dwarf is used to phase the $I_C JH$ light curves shown in the left-column panels.  The amplitude of this variation increases toward shorter wavelengths. The actual observations are repeated over each phase.  The right-column panels are phased to the secondary's rotation period of 14.05~d; interestingly, its amplitudes decreases toward shorter wavelengths.  Superimposed in each panel is a sinusoid fit representing the modulation due to the rotation of each component as described in \S\ref{spots}.}
\end{center}
\end{figure}

We begin by determining the properties of spots on the primary required to
reproduce the low-amplitude, periodic variability observed in the light curves.
The primary's variability amplitudes were measured by fitting a sum of two 
sinusoids to the out-of-eclipse data in each of the $I_C JH$ bands, 
one sinusoid corresponding to the rotation period of the primary
at 3.293~d and another for the secondary at 14.05~d (Fig.~\ref{fig-phased-lc}).
The amplitudes of the 3.29-d signal were then scaled up by the components' relative 
luminosities, since the observed amplitudes are diluted by
the light from the secondary. 

\begin{figure}
\begin{center}
\includegraphics[angle=90,width=6.5in]{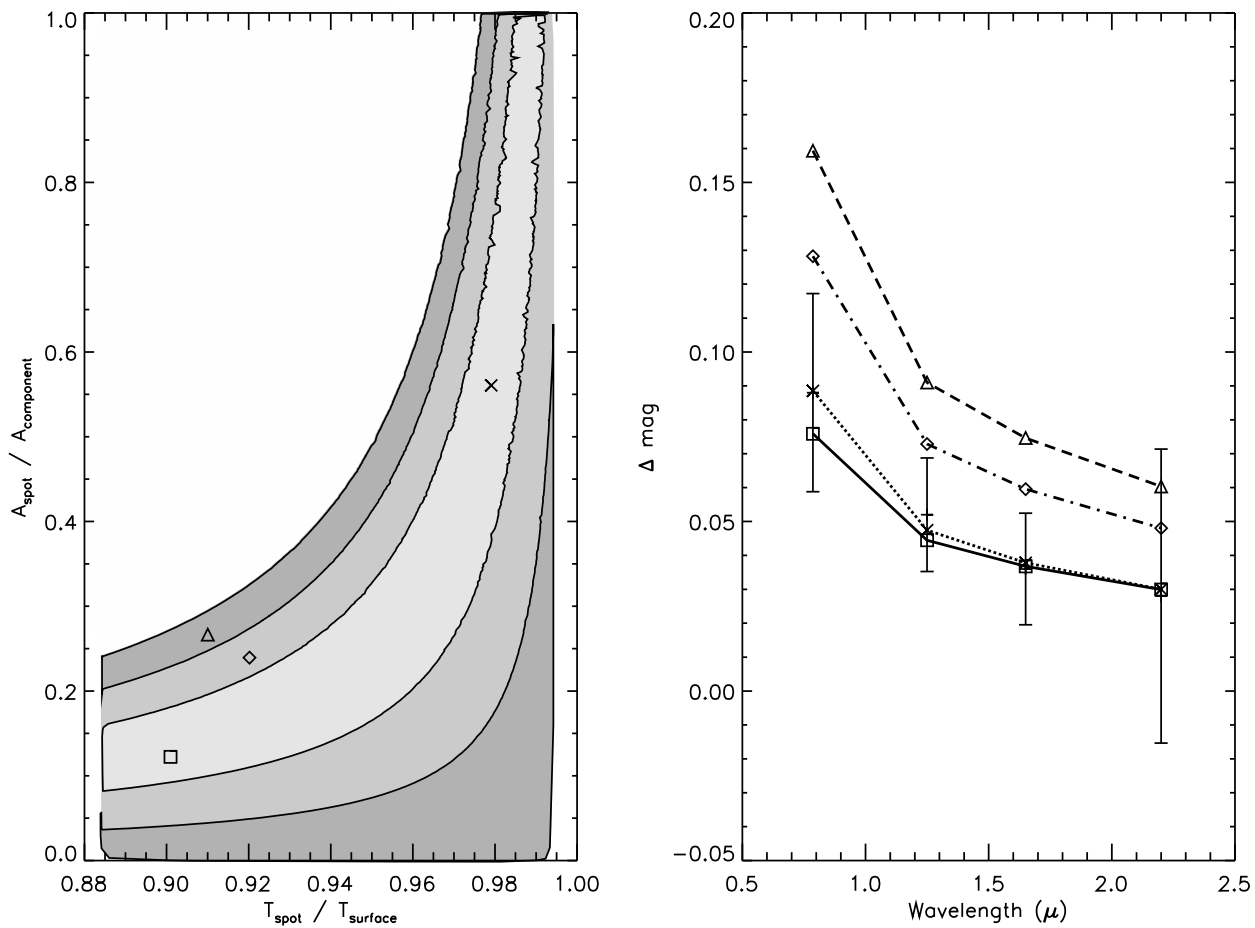}
\caption{\label{fig-amps}
Modeling of the Wavelength-Dependent Photometric Variability Using an Analytical Inversion Technique.  Using this technique, based on a two-component blackbody radiation \citep{bouvier}, to fit the measured low-amplitude photometric variability of \sm's light curves, we can estimate the spot temperature relative to the photosphere and the effective areal coverage of the spots. Because of the inherent degeneracy of spot modeling, a change in the temperature ratio maybe counteracted with an appropriate change in the areal coverage; the inversion technique renders not a single spot configuration but a family of solutions that describe the observed variability.
In the left-hand panel, the central region of the contours corresponds to those solutions for which the analytical amplitudes ($\Delta$ mag) fall within the 1-$\sigma$ photometric uncertainties of all of the observed bands.  The second level of contours represents the solutions that fall within the 2-$\sigma$ photometric uncertainties and the third those that fall within the 3-$\sigma$ uncertainties.  The right-hand panel shows the observed amplitudes of the photometric variation at the different wavelengths with 1-$\sigma$ error bars; for comparison the modeled amplitudes corresponding to the spot parameters marked by the four points in the left-hand panel are overplotted.  
The cross-point and the dotted line correspond to the fit with lowest $\chi^2$; the square-point and the continuous line are representative of the parameters chosen for the treatment of spots in the subsequent light curve modeling; the diamond-point and the dot-dash line denote a point on the second level contours, and the triangle-point and the dashed line correspond to a solution on the third contour level.
}
\end{center}
\end{figure}

These amplitudes were fit using an analytic model
based on a two-component blackbody as described by \citet{bouvier}. 
The free parameters are the spot temperature relative to the
photosphere and the spot areal coverage. The areal coverage 
parameter is an ``effective" area, i.e., it is really a measure of the 
ratio in spot coverage between the least and most spotted faces
of the surface and is thus a measure of the degree of spot asymmetry.
The results of this first-order analysis of the spot parameters are shown
in Fig.~\ref{fig-amps}.  A family of solutions is found, such
that a change in the spot temperature factor may be counterbalanced by a
change in the areal coverage.  
As one example, the observed light-curve variations can be fit with a
spot that is $\sim 10$\% cooler than the photosphere and that has an effective
areal coverage of $\sim 10$\%.
For the purposes of our modeling, and for simplicity, we placed a small cool spot 
with this temperature and area at the equator of the primary and allowed 
{\tt PHOEBE} to adjust the spot's longitude to match the phasing
of the observed variations (Fig.~\ref{fig-phased-lc}). 
We emphasize that the spot parameters are degenerate and we do not claim that the
adopted parameters are accurate in an absolute sense. Rather, they should be taken
as representative of the asymmetric component of the primary's spot distribution
that causes the observed low-amplitude variability modulated on the primary's
3.29-d rotation period (Fig.~\ref{fig-phased-lc}).

We ran a new light curve solution with {\tt PHOEBE}, this time including
the small spot on the primary as above, in order to check the influence of the
spot specifically on the derived temperature ratio. The best-fit system 
parameters are changed insignificantly. The temperature ratio in particular is 
changed from the value in Table~\ref{table-params} by less than $1 \sigma$. 
This is not surprising given the small areal coverage and temperature 
contrast of the spot and considering that in \pii\ we obtained a nearly identical
temperature ratio with a purely spotless model.
The inclusion of a small spot on the primary as required to fit the observed
low-amplitude variability is by itself not sufficient to explain the observed
temperature reversal in the system.

Therefore we next added a large cool spot at the pole of the primary. 
Assuming that the rotational and orbital axes of the system are parallel, and since 
$i\approx 90^\circ$, the effective areal coverage of a polar spot will not change
with rotational phase as seen by the observer. Thus this polar spot represents 
the {\it symmetric} component of the primary's spot distribution that, if it
covers a sufficiently large area, may cause an overall suppression of the primary's 
effective temperature
without producing additional variations with rotational 
phase\footnote{In fact, even a polar spot will cause a small variation 
{\it during eclipse}, however this effect is
$\sim 0.05$\% in the \ic\nir\ bands for the adopted spot parameters,
and is thus below the threshold of detectability given our 
photometric precision of $\sim 1$\%.}.
The evolutionary models of \citet{bar98} predict an effective 
temperature of 2880~K for a brown dwarf with the mass of the \sm\ 
primary at an age of 1~Myr, so we set the photospheric temperature 
of the primary to this value and re-fit the light curves with {\tt PHOEBE}, this 
time including both a small equatorial spot as before together with a large polar 
spot as described above, both with temperatures 10\% cooler than the photosphere.
The areal coverages of the two spots were left as free parameters,
and attained best-fit values of 8\% and 65\%, respectively.

Finally, we added a small equatorial spot on the secondary, again with a 
temperature 10\% cooler than the photosphere, representing the surface inhomogeneity
that produces the observed variations modulated on the secondary's 14.05-d rotation 
period (Fig.~\ref{fig-phased-lc}). Using {\tt PHOEBE} we performed a final 
simultaneous fit for the sizes of the spots on both the primary and secondary.
The final best-fit spot areal coverage factors for the smal spot on the
primary, the small spot on the secondary, and the large spot on the primary 
were 7\%, 3\%, and 65\%, respectively.


In reality, the observed variability of the magnetically inactive secondary 
is not likely to be caused by a magnetic spot, but perhaps more likely by 
dust in its atmosphere (\S\ref{period}). Our
light-curve modeling code does not currently incorporate a physical treatment 
of such dust features, and thus we use the spot modeling capability as a surrogate. 
In addition, we found that there is a near-total degeneracy between
the sizes of the small spots on the primary and secondary if their
temperatures are left as free parameters. That is, in the same way that
the temperature and size of an individual spot are degenerate (see
Fig.~\ref{fig-amps}), the sizes of the two spots relative to one another 
are degenerate unless their temperatures are fixed. Thus we have taken the
simplifying approach of fixing the spot temperatures to be 10\% cooler
than the surrounding photosphere on both the primary and secondary. 
The spot sizes are then constrained by the observed variability amplitudes
(Fig.~\ref{fig-phased-lc}).
Similarly, we have chosen not to include a large polar spot on the secondary as
we have on the primary.
The spot areas that we quote above are the formal best-fit 
values, however we caution that the properties we have determined for 
the inhomogeneity on the secondary should be taken as qualitative.
More important for our analysis here, the properties of the spots on the 
magnetically active primary are minimally affected by the presence of the
low-amplitude variability from the secondary, regardless of its true nature.

\begin{figure}
\begin{center}
\includegraphics{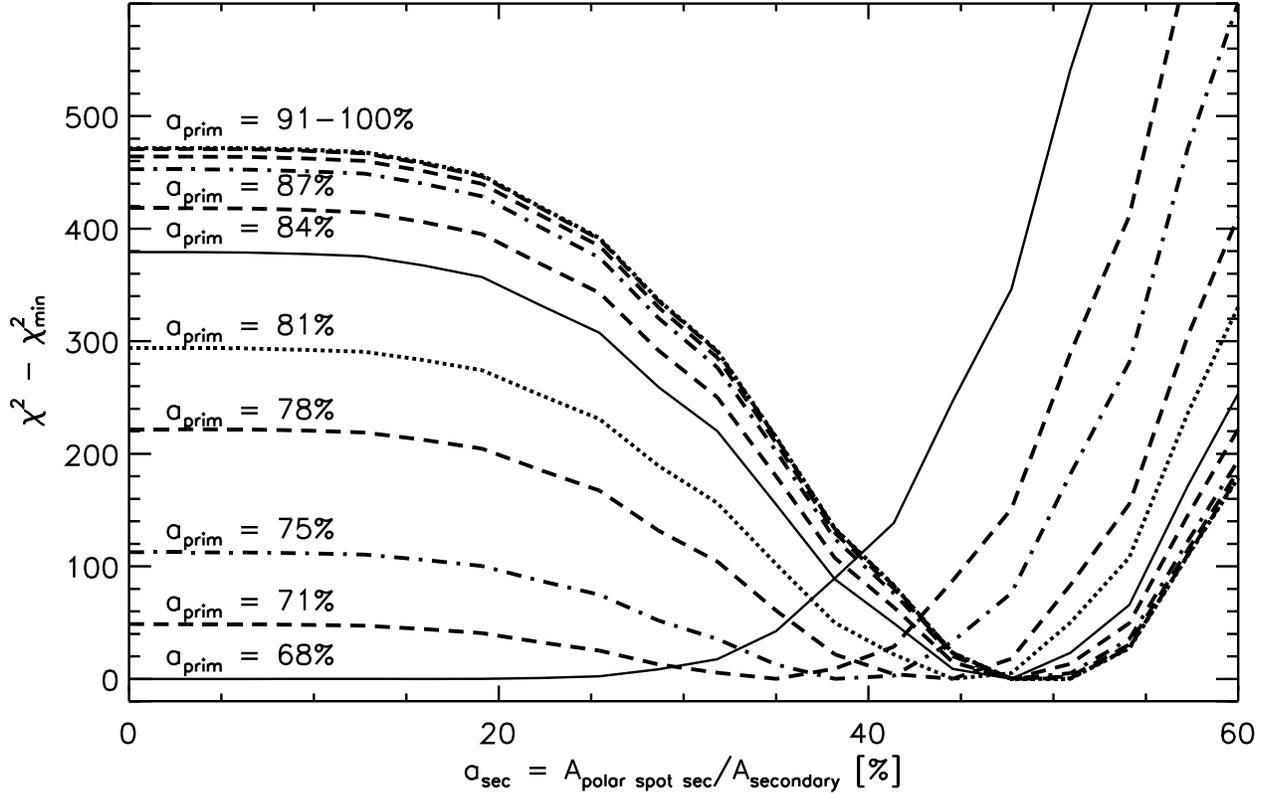}
\caption{\label{fig-secspots}
Degeneracy Between Large Polar Spots on the Primary and the Secondary Components of \sm.  
This figure shows the $\chi^2$ convergence of the light curve fitting of \sm\
when large polar spots are included on both of the components, with the temperatures
of the components fixed at the values predicted by the theoretical evolutionary models
of \citet{bar98}.
If a polar spot is included only on the primary, its areal coverage is required to be
65\% (see \S\ref{spots}). If a polar spot is also included on the secondary, the areal
coverage of the primary's polar spot must be increased to maintain the required temperature
ratio of the components.
All of the solutions shown in the figure are equivalent in terms of $\chi^2$ and
thus produce equally good fits to the data.
}
\end{center}
\end{figure}

Finally, we have not included a polar spot on the secondary, although 
from the standpoint of the light curve modeling alone it is possible to achieve equivalent
goodness-of-fit with polar spots on both components if their relative areal coverages
are adjusted so as to preserve the adopted photospheric temperature ratio (see 
Fig.~\ref{fig-secspots}).  We have taken the simplifying approach of including a polar
spot on the primary only because (1) the evidence suggests that the primary is the more
magnetically active of the two components \citep{reiners}, (2) the secondary's 
temperature is already in good agreement with the predictions of theoretical models
(\pii) and thus does not need to be suppressed by a large spot, and (3) as discussed above, 
the secondary's variability amplitudes do not indicate that it possesses magnetic spots.

\begin{figure}
\begin{center}
\includegraphics[width=6.5in]{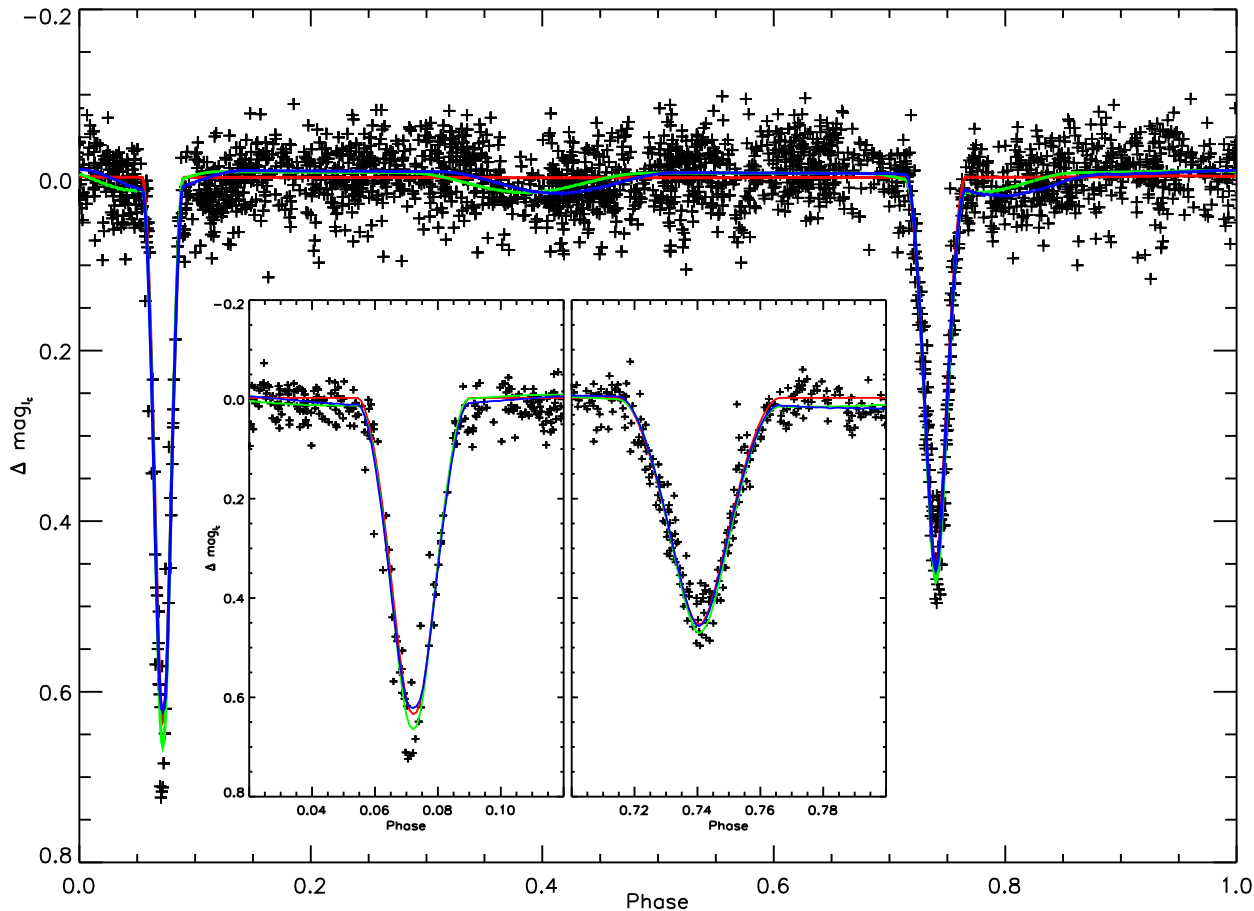}
\caption{\label{fig-lcfinal}
Light Curve Modeling Including Treatment of Spots.  The observed \ic\
light curve is plotted superimposed with three different synthetic models
each corresponding to a single orbital period starting at the time of
periastron: the red curve represents the spotless model; the green curve is
for the model where only the primary brown dwarf is spotted, and the blue
curve describes one in which both components have spots.  Both spotted
models are proposed to have spots that are 10\% cooler than the
surrounding photosphere and have an asymmetric constituent that describes
the low-amplitude photometric variability, by the use of a small, equatorial
spot on one or both of the components, and a symmetric constituent in order
to reconcile the more massive brown dwarf's effective temperature with
that expected from the evolutionary models \citep[e.g.][]{dant}, described
by polar spots with a large areal coverage.  The green model, with spots
only on the primary, has an equatorial spot that covers 10\% of the
surface (See Fig.~\ref{fig-amps}), and a polar spot with an areal coverage
of 65\%.  In the case where both components have spots (blue curve),
the low amplitude variability is described by an equatorial spot on the
primary that covers 7\% of its surface and another on the secondary with an
areal coverage of 3\%.  
Note that the non-synchronicity of the rotation with the orbital
period causes the effect of the spots on the light curve to change in phase
over time, rendering them noncoherent in the phase-folded data. 
}
\end{center}
\end{figure}

Fig.~\ref{fig-lcfinal} presents a comparison of the spotted and unspotted 
light curve models for the \ic\ band, the band in which the spot effects are 
most pronounced.
The synthetic light curves shown have been calculated over a single orbital period.
In view of the fact that the components do not rotate synchronously with one
another or with the orbital period, the effects of the spots on the light curves 
(such as the dip in the model at a phase of $\sim$0.4) will shift in orbital 
phase from one orbit to the next, and thus these variations are averaged out 
in the observed light curve which is phased over many orbital periods. 
We furthermore verified that the effects of the spots on the radial velocities
are negligible and thus do not affect any of the system's physical parameters 
that are determined kinematically (e.g.\ the masses).

The primary conclusion to be drawn from the above is that
the light curves of \sm\ can be well modeled by having the
primary component's photospheric temperature at the theoretically expected
value if $\sim 65$\% of its surface is covered with cool spots in a roughly
symmetric distribution. 
The small spot in our model represents the $\sim 10$\% asymmetry in the spot 
distribution that produces the observed low-amplitude periodic variations.

\section{Discussion\label{discussion}}

In order to simultaneously explain the observed low-amplitude variations and
the anomalously low effective temperature of the primary (more massive) component
in \sm, we have produced a model that includes a simple spot configuration 
of a small equatorial spot together with a very large polar spot.
The former represents the asymmetric component of the spot distribution
that produces the low-amplitude variations modulated on the primary's rotation
period, while the latter represents the symmetric component of the spot distribution
that causes an overall suppression of the effective temperature below its 
theoretically predicted value. In this model, the unspotted regions of the 
primary's surface have the theoretically predicted value of 2880~K \citep{bar98}.

The true distribution of spots on the primary's surface is probably more complex.
For example, 
a more realistic spot configuration might be one that resembles Jupiter's bands.  
In that case, a symmetric equatorial band with a temperature 10\% cooler than the
photosphere and extending $40^\circ$ above and below the equator would 
reproduce similarly the effect of the polar spot. 
The same result could be obtained by a leopard-print
pattern as that described by \citet{linnell} with an equivalent areal
coverage and equal spot temperature factor as the polar spot we modeled.
Either of these might describe
more accurately a physical configuration of spots for the primary.
Without direct Doppler imaging of \sm, it is not possible to 
more accurately pinpoint the true spot properties.  

We emphasize that there is nothing in our treatment of spots that prefers 
the primary's effective temperature to be 2880~K as dictated by the evolutionary 
models.  We could have chosen any other effective temperature for the photosphere 
surrounding the spots and achieved an equally good fit of the light curves 
by adjusting the spot temperature and/or areal coverage to compensate. Thus our
adoption of a primary effective temperature of 2880~K in the light curve solution 
of Fig.~\ref{fig-lcfinal} should
not be interpreted as a verification of the theoretical models.
In addition, it should be noted that in our model 
the {\it overall} surface brightnesses of the components
(integrating over both spotted and unspotted surface regions) are unchanged, such
that the primary's overall effective temperature is still lower than that of
the secondary.  This is an unavoidable consequence of the observed eclipse
depths, which ultimately require the secondary to be {\it effectively}
hotter than the primary.  The luminosities of the brown dwarfs thus also
remain the same regardless of the chosen effective temperature and 
corresponding spot configuration, because the overall surface brightnesses 
and radii are unaltered by our spot treatment.

Moreover, our modeling of spots treats only the radiative behavior of the surfaces 
of the brown dwarfs, not their underlying structure. Consequently our modeling does 
not serve as a detailed test of any structural or evolutionary effects 
caused by the surface magnetism that is likely responsible for the spots that
we have modeled. For example, \citet{cha07} have proposed that the temperature
reversal in \sm\ could be explained by a magnetically active primary with a
spot covering fraction of 50\% together with surface convection that has been
magnetically suppressed to a very low $\alpha = 0.5$ \citep[as opposed to the usual
$\alpha\approx$~1--2; e.g.][]{stass04}. They also suggest that suppressed convection 
may explain why the measured radius of the primary is $\sim 10$\% larger
than predicted by their theoretical mass-radius relationship.  Their exploratory 
treatment assumes ``black" (i.e.\ 0~K) spots, whereas our modeled spots have a more
physically realistic temperature 10\% cooler than the photosphere, so the total
spot-covering fraction of $\sim 75$\% that we find for the primary 
(large polar spot plus small equatorial spot) may in fact be consistent with 
the $\sim 50$\% coverage adopted by \citet{cha07}. In addition, we have 
empirically determined the radii of \sm\ with an accuracy of $\sim 1$\%, however 
our light curve modeling cannot confirm whether the radii have been altered
in some way by the presence of spots or by magnetically suppressed convection.



\section{Summary\label{summary}}

As the first known eclipsing binary where both components are brown dwarfs,
\sm\ is a paramount example against which theoretical brown dwarf formation
and evolutionary models and other low-mass objects will be compared.
\citet{nt06} and \citet{pii} established the young and low-mass nature of the 
binary, and moreover identified a surprising reversal of temperatures in which 
the primary (more massive) brown dwarf is cooler than the secondary.
Here, we reanalyze the previously published radial velocities and \ic-band
light curve together with newly obtained \nir\ light curves. We confirm the
surprising temperature reversal. In addition, our analysis improves the 
measurement of the system's parameters and permits a detailed modeling of magnetic 
spots on the brown dwarfs that may be altering their surface properties.

The masses of the components priorly reported to have 
uncertainties of $\sim$10\% (\pii) have been here determined with an
accuracy of $\sim$6.5\%, and the radii with an accuracy of $\sim$1.5\%. 
In addition, through a detailed analysis of the variability observed in
the light curves out of eclipse, the rotation periods of both brown dwarfs are 
measured to be $P_{\rm rot,1} = 3.293$~d and $P_{\rm rot,2} = 14.05$~d. Thus
the brown dwarfs rotate non-synchronously relative to the orbital motion and
relative to one another, perhaps due to the youth of the system
\citep[$\sim 1$ Myr;][]{nt06,pii}.
These rotation periods are in agreement with those expected from the radii and
the spectroscopically measured $v \sin{i}$. 

\citet{reiners} have suggested that the rapid rotation of the primary
brown dwarf, together with its strong $H\alpha$ emission, implies that it
is strongly magnetically active. In addition, \citet{cha07} have proposed that
a strong magnetic field on the primary could produce cool spots covering a
large fraction of its surface, thereby suppressing its effective temperature
and thus explaining why its temperature is lower than expected and apparently 
lower than the secondary's.

In this paper we have demonstrated that a detailed spectro-photometric 
modeling of \sm\ including the treatment of spots is
consistent with these ideas and in particular is able to 
resolve the apparent reversal of the temperature ratio.  
In order to reconcile the observed effective temperatures with those
predicted by theoretical models, the primary brown dwarf must be heavily spotted.
This `spottedness' must be more or less symmetric to agree with the 
low-amplitude variability observed in the light curves, and it must have large
effective areal coverage.  Thus we modeled a two-spot configuration on
the primary's surface: a large polar spot with an areal coverage of $\sim$65\% 
to account for the lower-than-expected surface brightness, 
and an equatorial spot covering $\sim$10\% of the surface for the purpose of 
introducing the asymmetry responsible for the observed
low-amplitude photometric variability modulated on the rotation period. 
With this configuration, we are able to successfully reproduce the observed 
light curves with the primary having an effective temperature at the theoretically
predicted value.
Other geometries for the spot configuration---such as an equatorial band akin 
to those on Jupiter---would achieve the same effect. 

To be clear, from the standpoint of the light-curve modeling alone there is no
need for a large spot-covering fraction on either brown dwarf. 
A small areal coverage of $\sim$10\% is sufficient to
model the low-amplitude variations that we observe in the light curves. Our
aim here has been to demonstrate as proof of concept that the spots on the primary 
are capable of explaining its suppressed effective temperature in a manner that is
consistent both with recent empirical findings of enhanced activity on the
primary \citep{reiners} and recent theoretical results on the effects of
such activity on the physical properties of young brown dwarfs \citep{cha07}.

\acknowledgments
We are grateful to Guillermo Torres for painstakingly reviewing our calculations of system parameters and for helpful discussions on the determination of parameter uncertainties.  
We acknowledge funding support to Y.G.M.C.\ by a CONACYT fellowship from M\'exico.
K.G.S.\ acknowledges the support of NSF Career grant AST-0349075 and a Cottrell
Scholar award from the Research Corporation.
\bibliographystyle{apsrmp}
\bibliography{sm4147JHK-090511}

\begin{deluxetable}{crcrcc}
\tablecolumns{6}
\tablewidth{0pc}
\tablecaption{Photometric Time Series Observations of \sm\label{table-seasons}}
\tablehead{
\colhead{\phn} &
\colhead{UT Dates}    & 
\colhead{Julian Dates Range} &
\colhead{Filter}& 
\colhead{Exp\tablenotemark{1}} &
\colhead{Obs\tablenotemark{2}}}

\startdata

I & 2003 10 09 -- 2004 03 16 & 
2452922.728 -- 2453081.568 & $H$\phn & 525 & 303 \\

II & 2004 10 01 -- 2004 11 30 & 
2453280.731 -- 2453340.726 & $J$\phn & 490 & 105\\
\phn& \phn & 
2453280.736 -- 2453340.733 & $K_{s}$ & 490 & 104 \\

III & 2005 02 01 -- 2005 03 15 & 
2453403.540 -- 2453445.589  & $J$\phn & 490 & 123 \\ 
\phn & \phn & 
2453403.532 -- 2453445.579 & $H$\phn & 490 & 123\\
\phn & \phn & 
2453403.547 -- 2453445.595  & $K_{s}$ & 490 & 115\\

IV &2005 10 02 -- 2005 12 23 & 
2453646.828 -- 2453728.701  & $J$\phn & 490 & 55 \\ 
\phn & \phn & 
2453646.821 -- 2453728.694 & $H$\phn & 490 & 53\\
\phn & \phn & 
2453646.836 -- 2453728.717 & $K_{s}$ & 490 & 103\\

V & 2006 01 09 -- 2006 04 09 & 
2453745.651 -- 2453835.506  & $J$\phn & 490 & 81 \\ 
\phn & \phn & 
2453745.643 -- 2453835.498 & $H$\phn & 490 & 89\\
\phn & \phn & 
2453745.658 -- 2453835.514 & $K_{s}$ & 490 & 64\\

\enddata
\tablenotetext{1}{\footnotesize Total exposure time in seconds of the seven dithered positions.} 
\tablenotetext{2}{\footnotesize Number of observations per season.}
\end{deluxetable}

\begin{deluxetable}{lr}
\tablecaption{Differential $J$ band Light Curve of \sm\label{table-j}}
\tablewidth{0pc}
\tabletypesize{\small}
\tablehead{\colhead{HJD\tablenotemark{a}} & \colhead{$\Delta$$m$\tablenotemark{b}} }
\startdata
2453311.723468 & -0.02137 \\
2453321.645380 & 0.00305 \\
2453327.667177 & 0.09047 \\
2453327.736855 & 0.25355 \\
2453337.627205 & 0.44196 \\
2453337.712161 & 0.30654 \\
2453340.661636 & 0.02005 \\
2453291.837179 & 0.06698 \\
2453301.833196 & 0.13945 \\
2453280.731279 & -0.01333 \\
2453280.795035 & 0.01343 \\
2453280.850294 & -0.02312 \\
2453281.725007 & 0.00358 \\
2453281.790626 & 0.01219 \\
2453281.842875 & 0.00321 \\
\enddata
\tablenotetext{a}{\footnotesize Heliocentric Julian Date}
\tablenotetext{b}{\footnotesize Differential $J$ magnitude}
\tablecomments{\footnotesize This is only a portion of the complete table.}
\end{deluxetable}

\begin{deluxetable}{lr}
\tablecaption{Differential $H$ band Light Curve of \sm\label{table-h}}
\tablewidth{0pc}
\tabletypesize{\small}
\tablehead{\colhead{HJD\tablenotemark{a}} & \colhead{$\Delta$$m$\tablenotemark{b}} }
\startdata
2453426.520578 & 0.03610 \\
2453445.485528 & 0.01546 \\
2453428.650735 & -0.00740 \\
2453415.616726 & 0.00619 \\
2453415.677288 & 0.06725 \\
2453425.564072 & 0.37523 \\
2453425.630860 & 0.53849 \\
2453425.662732 & 0.47590 \\
2453406.539051 & 0.01280 \\
2453409.535701 & -0.01483 \\
2453409.619548 & 0.00244 \\
2453435.510760 & 0.27719 \\
2453435.569216 & 0.11036 \\
2453435.621445 & 0.02131 \\
2453436.510477 & 0.00301 \\
\enddata
\tablenotetext{a}{\footnotesize Heliocentric Julian Date}
\tablenotetext{b}{\footnotesize Differential $H$ magnitude}
\tablecomments{\footnotesize This is only a portion of the complete table.}
\end{deluxetable}

\begin{deluxetable}{lr}
\tablecaption{Differential $K_S$ band Light Curve of \sm\label{table-k}}
\tablewidth{0pc}
\tabletypesize{\small}
\tablehead{\colhead{HJD\tablenotemark{a}} & \colhead{$\Delta$$m$\tablenotemark{b}} }
\startdata
2453336.725515 & -0.07627 \\
2453336.758641 & -0.04853 \\
2453428.666590 & -0.04077 \\
2453415.586925 & -0.03317 \\
2453415.632014 & -0.07676 \\
2453415.692587 & -0.00632 \\
2453425.579580 & 0.42007 \\
2453425.646588 & 0.49068 \\
2453425.678437 & 0.44216 \\
2453409.550989 & 0.01162 \\
2453409.634848 & 0.00240 \\
2453435.526476 & 0.21726 \\
2453435.584515 & 0.11184 \\
2453435.637277 & 0.00116 \\
2453438.537777 & 0.08671 \\
\enddata
\tablenotetext{a}{\footnotesize Heliocentric Julian Date}
\tablenotetext{b}{\footnotesize Differential $K_S$ magnitude}
\tablecomments{\footnotesize This is only a portion of the complete table.}
\end{deluxetable}

\begin{deluxetable}{ccccc} 
\tablecaption{Periodicities Detected by Season and Passband\label{table-periods}}
\tablewidth{0pc}
\tabletypesize{\small}  
\tablehead{\colhead{Season\tablenotemark{a}} & \multicolumn{2}{c}{$P_{\rm rot,1} = 3.29$~d} & \multicolumn{2}{c}{$P_{\rm rot,2} = 14.05$~d}\\ \phn & \ic & $JH$ & \ic & $JH$} 
\startdata
I\tablenotemark{b} & 	\checkmark & \checkmark & \nodata & \checkmark \\
II\tablenotemark{b} & 	\checkmark & \checkmark & \checkmark & \checkmark \\ 
III &			\checkmark & \checkmark & \nodata & \checkmark \\       
IV &			\checkmark & \checkmark & \nodata & \checkmark \\       
V &			\checkmark & \checkmark & \checkmark& \nodata \\       
\enddata
\tablenotetext{a}{\footnotesize See Table~\ref{table-seasons} for details of the 
observing campaigns.}
\tablenotetext{b}{\footnotesize Only $J$ or $H$ were observed during this season.}
\end{deluxetable}

\begin{deluxetable}{lrclcrcl}
\tabletypesize{\small}
\tablecolumns{8}
\tablewidth{0pc}
\tablecaption{Orbital and Physical Parameters of \sm\label{table-params}}
\tablehead{
\colhead{\phn}&
\multicolumn{3}{c}{RVs $+$ \ic \tablenotemark{1}} & \colhead{\phn}&
\multicolumn{3}{c}{RVs $+$ \nir\ $+$ \ic} }

\startdata
Orbital period, $P_o$ (days)              &      \multicolumn{7}{c}{9.779556 $\pm$ 0.000019} \\
Time of periastron, $T_o$ (Besselian year)
			&2001.863765&$\pm$&0.000071&
		\phn	&2001.8637403&$\pm$&0.0000062\\
Eccentricity, {\it e}   &0.3276&$\pm$&0.0033&
                \phn    &0.3216&$\pm$&0.0019\\
Orientation  of periastron, $\omega$ ( \degr)
                         &217.0&$\pm$&0.9&
                \phn    &215.3&$\pm$&0.5\\
Semi-major axis, $a\sin{i}$ (AU)
                        &0.0406&$\pm$&0.0010&
                \phn    &0.0407&$\pm$&0.0008 \tablenotemark{\dagger}\\
Inclination angle, $i$ ( \degr)
                        &89.2&$\pm$&0.2&
                \phn    &88.49&$\pm$&0.06\\
Sytemic velocity, $v_\gamma$ (km s$^{-1}$)
                        &24.1&$\pm$&0.4&
                \phn    &24.1&$\pm$&0.3 \tablenotemark{\dagger}\\
Primary semiamplitude, $K_1$ (km s$^{-1}$)
                        &18.49&$\pm$&0.67&
                \phn    &18.61&$\pm$&0.55\\
Secondary semiamplitude, $K_2$ (km s$^{-1}$)
                        &29.30&$\pm$&0.81&
                \phn    &29.14&$\pm$&1.40\\
Mas ratio, $q \equiv M_2/M_1$
                        &0.631&$\pm$&0.015&
                \phn    &0.639&$\pm$&0.024 \tablenotemark{\dagger}\\
Total mass, $M\sin^3{i}$ (M$_\sun$)
                        &0.0932&$\pm$&0.0073&
                \phn    &0.0936&$\pm$&0.0051 \tablenotemark{\dagger}\\
Primary mass, $M_1$ (M$_\sun$)
                        &0.0572&$\pm$&0.0045&
                \phn    &0.0572&$\pm$&0.0033\\
Secondary mass, $M_2$ (M$_\sun$)
                        &0.0360&$\pm$&0.0028&
                \phn    &0.0366&$\pm$&0.0022\\
Primary radius, $R_1$ (R$_\sun$)
                        &0.675&$\pm$&0.023&
                \phn   &0.690&$\pm$&0.011\\
Secondary radius, $R_2$ (R$_\sun$)
                        &0.486&$\pm$&0.018&
                \phn   &0.540&$\pm$&0.009\\
Primary gravity, log $g_1$
                        &3.54&$\pm$&0.09&
                \phn    &3.52&$\pm$&0.03\\
Secondary gravity, log $g_2$
                        &3.62&$\pm$&0.10&
                \phn    &3.54&$\pm$&0.03\\
Primary surface potential, $\Omega_1$
                        &\multicolumn{3}{c}{\nodata}&
                \phn    &13.63&$\pm$&0.18 \\
Secondary surface potential, $\Omega_2$
                        &\multicolumn{3}{c}{\nodata}&
                \phn    &12.00&$\pm$&0.16 \\
Primary synchronicity parameter, $F_{1}$
			&\multicolumn{3}{c}{\nodata}&
                \phn    &2.9725&$\pm$&0.0009 \\
Secondary synchronicity parameter, $F_{2}$
                        &\multicolumn{3}{c}{\nodata}&
                \phn    &0.6985&$\pm$&0.0025 \\ 
Effective temperature ratio, $T_{\rm eff,2}/T_{\rm eff,1}$
                        &1.064&$\pm$&0.004&
                \phn    &1.050&$\pm$&0.004\\

\enddata
\tablenotetext{1}{\footnotesize Previously published results (\pii).}
\tablenotetext{\dagger}{\footnotesize  
The uncertainties in these parameters were conservatively estimated from the formal errors of a fit to the RV data alone. See \S 3.2. }
\end{deluxetable}

\end{document}